\RequirePackage{lineno}
\documentclass[a4paper,11pt]{article}

\usepackage{jheppub} 

\usepackage[T1]{fontenc} 

\newcommand{\Rdr}{R_{\Delta R}}
\newcommand{\Rdphi}{R_{\Delta \phi}}
\newcommand{\Rtt}{R_{3/2}}

\newcommand{\Dphimax}{\Delta \phi_{\rm max}}
\newcommand{\Dphi}{\Delta \phi_{\rm dijet}}

\newcommand{\ptmax}{p_{T}^{\rm max}}
\newcommand{\ptmin}{p_{T \rm min}}

\newcommand{\ystar}{y^*}
\newcommand{\ystarmax}{y^*_{\rm max}}
\newcommand{\yboost}{y_{\rm boost}}

\newcommand{\mur}{\mu_R}
\newcommand{\muf}{\mu_F}

\newcommand{\as}{\alpha_s}

\newcommand{\asmz}{\alpha_s(M_Z)}

\newcommand{\pythia}{{\sc pythia}}
\newcommand{\herwig}{{\sc herwig}}

\newcommand{\Rcone}{R_{\rm cone}}

\newcommand{\ord}{{\cal O}}
\newcommand{\ppbar}{p{\bar{p}}}

\title{\boldmath A new quantity for studies of dijet azimuthal decorrelations}

\author{M. Wobisch,}
\author{K. Chakravarthula,}
\author{R. Dhullipudi,}
\author{L. Sawyer,}
\author{M. Tamsett}

\affiliation{Department of Physics, Louisiana Tech University,\\
             600 West Arizona Ave., Ruston, USA}

\emailAdd{wobisch@latech.edu}
\emailAdd{kchakrav@fnal.gov}   
\emailAdd{ram.dhullipudi@cern.ch}
\emailAdd{sawyer@latech.edu}
\emailAdd{tamsett@cern.ch}

\abstract{We introduce a new measurable quantity, $\Rdphi$,
for studies of the rapidity and transverse momentum dependence 
of dijet azimuthal decorrelations in hadron-hadron collisions.
In pQCD, $\Rdphi$ is computed as a ratio of three-jet and 
dijet cross sections in which the parton distribution functions 
cancel to a large extent.
At the leading order, $\Rdphi$ is proportional to $\as$,
and the transverse momentum dependence of $\Rdphi$ can therefore be 
exploited to determine $\as$. 
We compute the NLO pQCD theory predictions and non-perturbative 
corrections for $\Rdphi$ at the LHC and the Tevatron
and investigate the corresponding uncertainties.
From this, we estimate the theory uncertainties for $\as$ determinations 
based on $\Rdphi$ at both colliders.
The potential of $\Rdphi$ measurements for tuning Monte Carlo 
event generators is also demonstrated.
}

\keywords{jets, hadronic colliders, QCD}

\begin{document} 

\noindent
\hfill \mbox{FERMILAB-PUB-12-618-E}


\maketitle
\flushbottom

\section{Introduction}
\label{sec:intro}

Theory predictions for inclusive jet and dijet cross sections 
in hadron-hadron collisions at fixed order in 
perturbative Quantum Chromodynamics (pQCD)
are currently available at next-to-leading order (NLO) 
in the strong coupling constant $\as$.
Using precise experimental data, these predictions have been well tested
and applied in determining the parton distribution functions (PDFs) 
of the proton, and $\as$~\cite{PDG2012}.
Direct tests of pQCD at higher orders require measurements of 
quantities probing multi-jet final states with three or more jets.
Quantities in which a cross section for the production of three or more jets
is normalized by a dijet cross section (or an inclusive jet cross section)
are ideal for $\as$ determinations.
These quantities are still sensitive to the degrees 
of freedom in the multi-jet final state and, in pQCD, 
proportional to (at least) $\ord(\as)$,
while the PDF sensitivity 
exhibited by a typical multi-jet cross section~\cite{Abazov:2011ub}
can be strongly reduced.
Examples of such quantities are the ratio of the inclusive three jet 
and dijet cross sections, 
$\Rtt$~\cite{Chatrchyan:2011wn,Aad:2011tqa,:2012sy},
and the average number of neighboring jets, $\Rdr$, which has recently 
been proposed, measured, and used to determine $\as$~\cite{:2012xi}.

\begin{figure}  
\centering
\includegraphics[width=3.5cm]{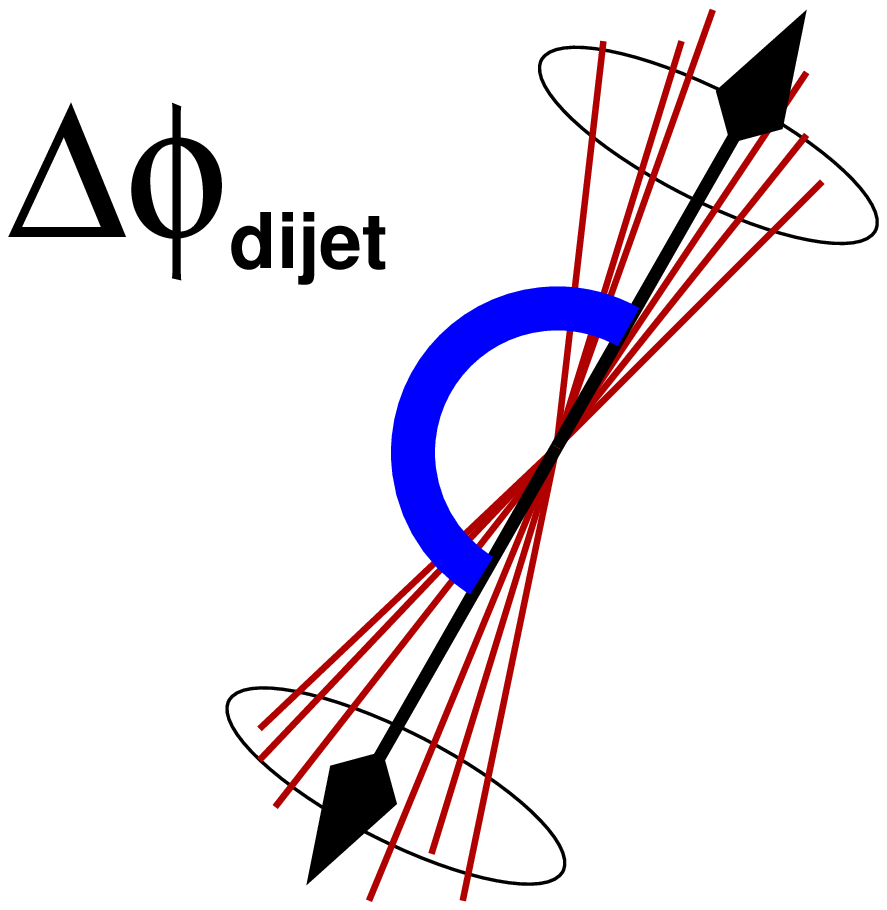} \, 
\includegraphics[width=3.5cm]{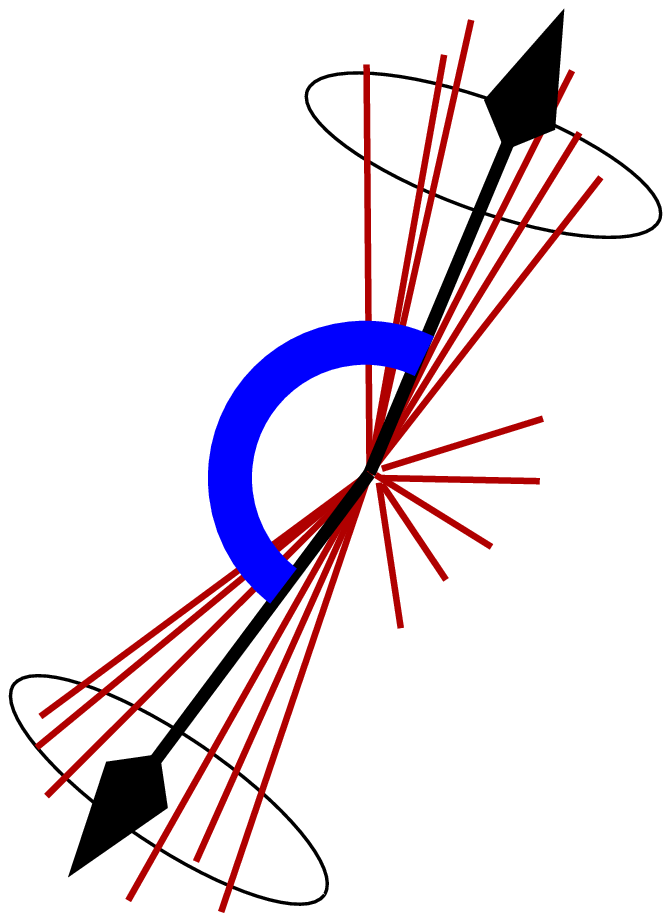} \, 
\includegraphics[width=3.5cm]{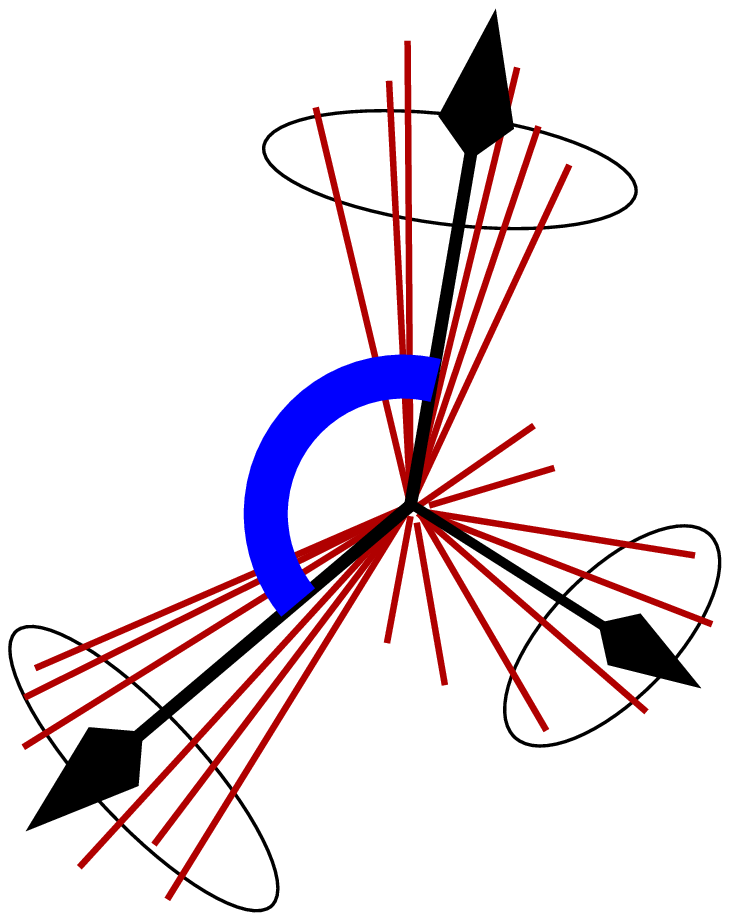} \,
\includegraphics[width=3.5cm]{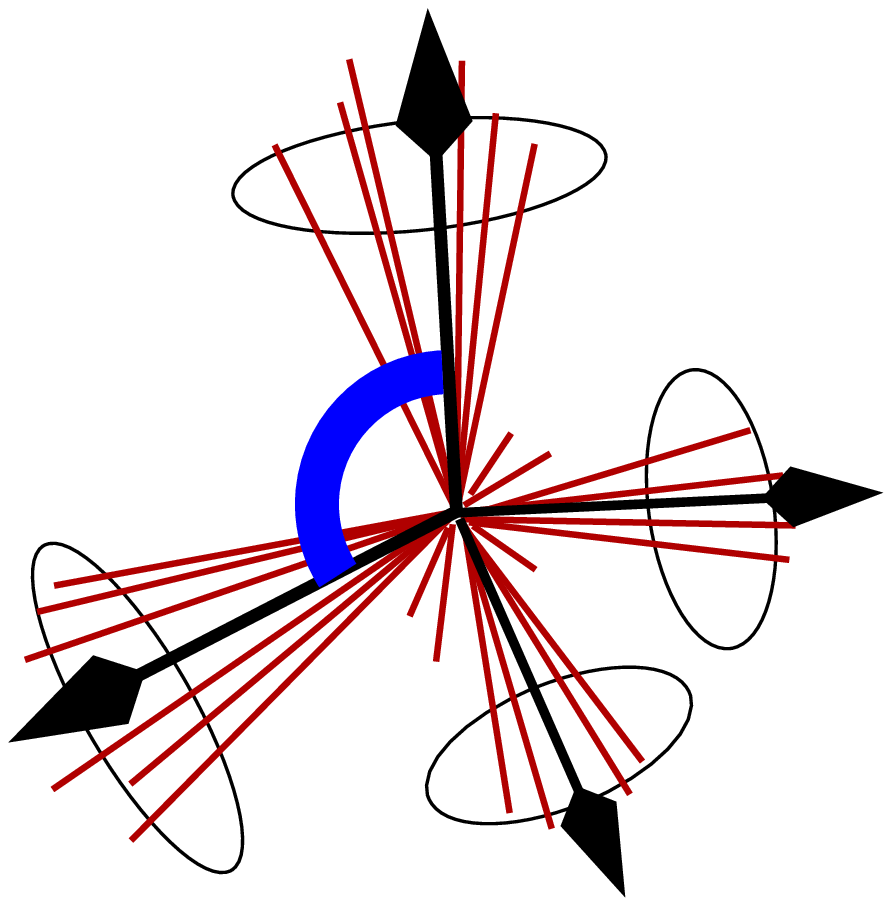}
\caption{\label{fig:dphi}
   A sketch of the angle $\Dphi$ in the azimuthal plane,
   in dijet events for different 
   amounts of additional radiation outside the dijet system.}
\end{figure}

A third related quantity is the dijet azimuthal decorrelation,
which studies the relative angle in the azimuthal plane between 
the two jets with the highest transverse momentum ($p_T$) 
$\Dphi = | \phi_{\rm jet1} - \phi_{\rm jet2} |$.
In calculations at $\ord(\as^2)$, dijet events have exactly two jets
with equal $p_T$, and their azimuthal angles     
are correlated such that $\Dphi = \pi$.
Deviations from $\Dphi = \pi$ (hereafter referred to as 
``azimuthal decorrelations'') are caused by additional radiation
which is not clustered into the two highest $p_T$ jets.
Additional radiation with small $p_T$ reduces $\Dphi$ by a small amount,
while high-$p_T$ radiation can reduce $\Dphi$ significantly
thereby 
leading to larger azimuthal decorrelations
as illustrated in figure~\ref{fig:dphi}.
Due to kinematic constraints, three-jet final states are restricted 
to $\Dphi > 2\pi/3$, while the phase space of $\Dphi < 2\pi/3$ 
is only accessible in final states with at least four jets.

The D\O\ collaboration has introduced the quantity
$(1/\sigma_{\rm dijet}) \cdot d \sigma_{\rm dijet} / d \Dphi$,
which is the dijet cross section differentially
in $\Dphi$, normalized by the inclusive dijet cross section 
$\sigma_{\rm dijet}$ (in the same kinematic range and
integrated over $\Dphi$)~\cite{Abazov:2004hm}.
This quantity was measured in $\ppbar$ collisions at $\sqrt{s}=1.96\,$TeV,
for different $\ptmax$ requirements, where $\ptmax$ is the highest jet $p_T$ 
in the event, and for a fixed $p_T$ requirement for the second leading $p_T$ 
jet.
For this quantity, the range from small to large azimuthal decorrelations
can be used to study the transition from soft to hard higher-order pQCD 
processes and the measurement results placed strong constraints on 
Monte Carlo parameters~\cite{begel}.
The same analysis strategy was later employed by the CMS and ATLAS 
collaborations using $pp$ collision data at $\sqrt{s}=7\,$TeV,
thus accessing larger $\ptmax$~\cite{Khachatryan:2011zj,daCosta:2011ni}.
The common approach focuses on the $\Dphi$ dependence; 
the $p_T$ dependence is not easily visible in these presentations.
Furthermore, in pQCD, dijet azimuthal decorrelations are predicted 
to depend not only on $p_T$, but also on the rapidities 
of the two leading $p_T$ jets.      
The measurements by the D\O, CMS, and ATLAS collaborations, however,
did not explore the rapidity dependence.

In this article, we propose a new quantity $\Rdphi$ for studying 
dijet azimuthal decorrelations with emphasis on the rapidity and 
the $p_T$ dependence.\footnote{Some initial studies for 
   experimental measurements of $\Rdphi$ have been made 
   in references~\cite{Dhullipudi,Chakravarthula}.}
The former aspect will allow us to perform novel tests of the pQCD predictions,
while the latter can be exploited for determinations of $\as$
and its running.
The article is structured as follows:
In section~\ref{sec:def} we motivate the variables used to study the 
rapidity and $p_T$ dependencies, and we give the definition
of $\Rdphi$.
In addition, we propose realistic scenarios for phase space regions 
in which $\Rdphi$ can be measured by the LHC and 
the Tevatron experiments.
Theory predictions for these scenarios are presented
in section~\ref{sec:thy}, including perturbative and non-perturbative
contributions.
The possible impact of $\Rdphi$ measurements on 
determinations of $\as$ and on Monte Carlo tuning
is discussed in section~\ref{sec:pheno}.

\section{Definition and Phase Space Scenarios \label{sec:def}}

The quantities $\Rtt$~\cite{Chatrchyan:2011wn,Aad:2011tqa,:2012sy} 
and $\Rdr$~\cite{:2012xi} are defined as ratios of multi-jet cross sections.
These ratios can be interpreted as the conditional probability
that an event with two high-$p_T$ jets also contains a third jet ($\Rtt$)
and as the average number of neighboring jets for a given jet ($\Rdr$).
We propose to study dijet azimuthal decorrelations,
using a quantity with a similar intuitive interpretation.
For this purpose we introduce the quantity $\Rdphi$.
Before we define $\Rdphi$, we motivate the variables used to study the 
rapidity and $p_T$ dependencies.
In addition, we propose realistic scenarios for measurements 
of $\Rdphi$ at the LHC and the Tevatron.

\subsection[Variables for the Rapidity and $p_T$ Dependence]%
{\boldmath Variables for the Rapidity and $p_T$ Dependence}
\label{subsec:var}

\begin{figure}  
\centering
\includegraphics[height=3.99cm]{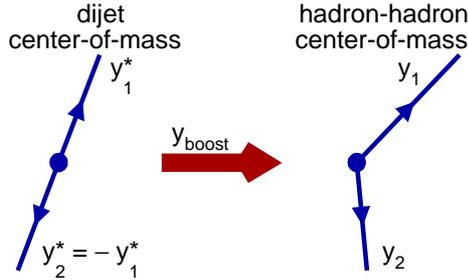} 
\caption{\label{fig:boost}
   Jet rapidity variables in the transverse plane
   in the dijet center-of-mass frame and
   in the hadron-hadron center-of-mass frame.
}
\end{figure}

\paragraph{The Rapidity Variable}

One of the main goals for the new quantity $\Rdphi$ is to measure
the rapidity dependence of dijet azimuthal decorrelations.
The previous analyses~\cite{Abazov:2004hm,Khachatryan:2011zj,daCosta:2011ni}
applied rapidity requirements for both jets in the hadron-hadron 
center-of-mass frame (i.e.\ the lab frame).
In general, this frame is, however, longitudinally boosted with respect 
to the center-of-mass frame of the hard subprocess 
(corresponding to $\yboost$), as shown in figure~\ref{fig:boost}.
In the approximation of $2 \rightarrow 2$ processes, 
the rapidities $y^*_1$ and $y^*_2$ (in the dijet center-of-mass frame)
have the same magnitude ($y^* \equiv |y^*_1| = |y^*_2|$), and are related 
to the rapidities $y_1$ and $y_2$ (in the hadron-hadron center-of-mass frame)
by
\begin{equation}
y_1  =  y_1^* + \yboost \qquad \hbox{and} \qquad
y_2  =  y_2^* + \yboost    \, .
\end{equation}
We propose to measure the rapidity dependence of $\Rdphi$ as a function 
of the variable $\ystar$ for a fixed requirement for the variable $\yboost$.
Both variables are given by
\begin{equation}
\yboost = (y_1 + y_2)/2 \qquad \hbox{and} \qquad
y^* = |y_1 - y_2|/2   \, ,
\end{equation}
where $y_1$ and $y_2$ are the respective rapidities of the two leading $p_T$ 
jets in the event.

\paragraph{\boldmath The $p_T$ Variable}

\begin{figure}  
\centering
\includegraphics[height=4.3cm]{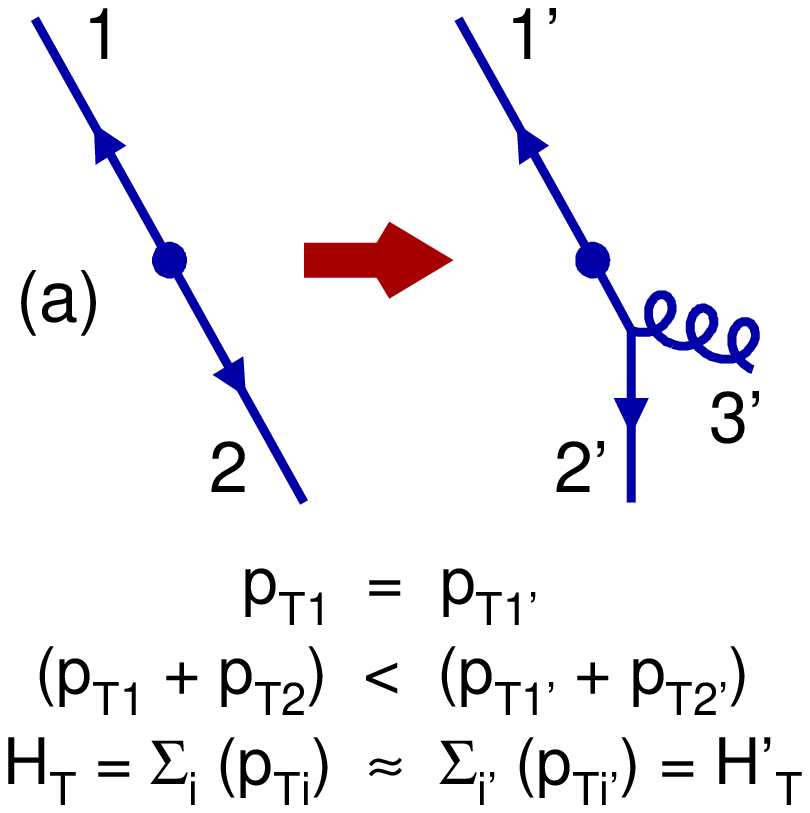} \hskip3cm
\includegraphics[height=4.3cm]{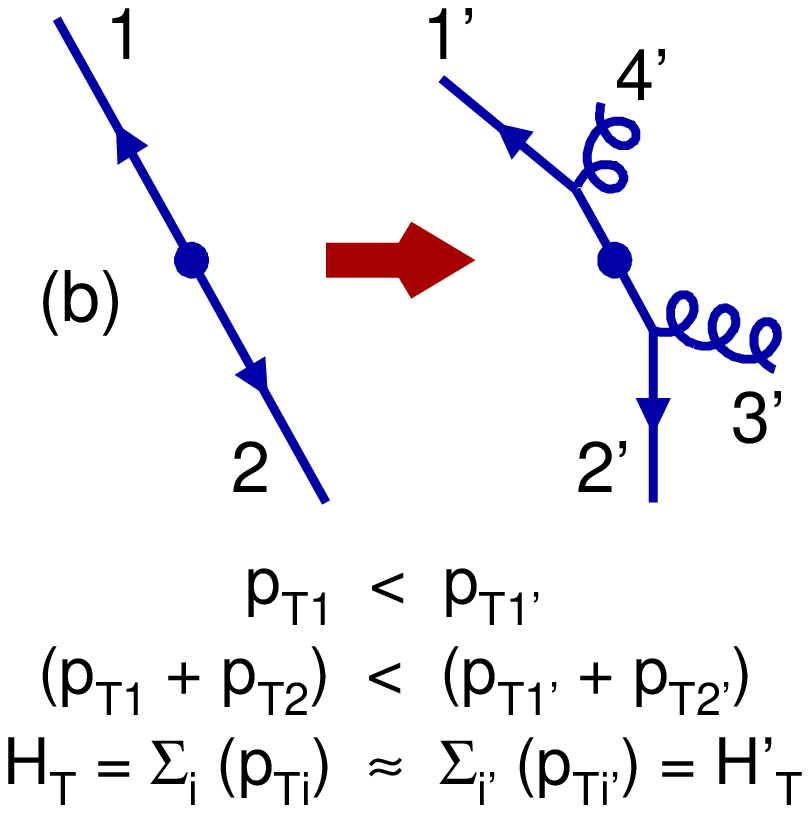} 
\caption{\label{fig:branching}
   Sketches of the azimuthal plane in which a three-parton (a) 
   and a four-parton final state (b)
   emerge from a two-parton final state through parton-branching,
   and the relations of different variables before and after the branching.
   The variable $H_T$ is approximately preserved in these processes.
}
\end{figure}

In the leading logarithmic approximation, an $n$-parton final state 
can be regarded as emerging from a two-parton final state 
through successive branching, as displayed in figure~\ref{fig:branching}
for three- and four-parton final states.
In this picture, a quantity such as $\Rtt$ may be interpreted as 
the branching probability from two to three final state partons
(for partons which have a sufficient angular separation 
to be resolved as individual jets, according to the jet definition).
However, this interpretation only holds if the quantity
is binned in a ``$p_T$-type'' variable that does not change
its value before and after the branching of the third parton.
Examples of such variables are the leading jet $p_T$, or, approximately, 
$H_T$ (defined as the scalar $p_T$ sum over all jets in an event).
A counter example is the variable $H_T^{(2)}=(p_{T1}+p_{T2})$,
defined as the scalar $p_T$ sum of the two leading jets, which is reduced after 
the branching displayed in figure~\ref{fig:branching}~(a).
For a quantity like the ratio of inclusive four-jet and dijet cross sections, 
$R_{4/2}$,  the leading jet $p_T$ could also be reduced,
e.g.\ by a branching as displayed in figure~\ref{fig:branching}~(b).

In the measurement of dijet azimuthal decorrelations, the angle $\Dphi$ 
is affected not only by the third jet but by all additional radiation in 
the event.
Therefore variables like $\ptmax$ or $H_T^{(2)}$ are not suited 
if we want to interpret the ratio $\Rdphi$ 
as the probability for parton branching.
Only the value of $H_T$ is approximately conserved after 
the branching processes.
With an ideal detector and in a clean environment, 
one might want to define $H_T$ as the scalar $p_T$ sum of all jets 
in the event, without any $p_T$ or $y$ requirements.
In practice, one has a limited detector $y$ acceptance, 
a limited knowledge of the detector response for low $p_T$ jets, 
plus contributions from the underlying event.
Therefore it is advisable to restrict the $p_T$ sum in the $H_T$ definition
to jets which are well measured, and for which non-perturbative
contributions are small, by requiring $p_{Ti} > \ptmin$.   
A limited detector $y$ acceptance can be taken into account
by requiring that the jets are contained inside
this acceptance region.
Since we study the rapidity dependence based on $\ystar$ which
is longitudinally boost invariant, 
we would like to preserve this property also for the $H_T$ definition.
Therefore the jet selection is not based on the absolute
jet rapidities $|y_i|$ in the lab frame, but on 
the longitudinally boost invariant quantity $|y_i - \yboost|$,
and $H_T$ is computed as
\begin{equation}
    H_T = \sum_{i \in C} p_{Ti}    \, ,
\label{eq:ht}
\end{equation}
based on all jets in the set $C$ which is defined as
\begin{equation}
 C = \{ \; i \; | \; 1 \le i \le n_{\rm jet} ; \; \mbox{and} \;
   p_{Ti} > p_{T \rm min} ; \; \mbox{and} \;
  |y_i - \yboost| < \ystarmax \}   \, .
\label{eq:}
\end{equation}
In this definition, $n_{\rm jet}$ is the total number of jets in the event,
and $\ptmin$ and $\ystarmax$ are parameters 
which can be chosen according to the experimental environment.
The value of $\ystarmax$ should be chosen at least as large as the maximum
accessible $y^*$ for two leading $p_T$ jets, to ensure that these 
are always members of the set $C$, and therefore included in the $H_T$ sum.

\subsection[Definition of $\Rdphi$]{\boldmath Definition of $\Rdphi$ 
\label{sec:defrdphi}}

With the criteria above, we propose to study dijet azimuthal 
decorrelations using the new quantity $\Rdphi$, which represents 
the fraction of all inclusive dijet events for which the two 
leading $p_T$ jets have a decorrelation of $\Dphi < \Dphimax$.
It is defined as
\begin{equation}
\Rdphi(H_T, y^*, \Dphimax) \, = \,
\frac{
\frac{d^2\sigma_{\rm dijet}(\Dphi < \Dphimax)}{dH_T \, dy^*} }%
{\frac{d^2\sigma_{\rm dijet}(\mbox{\footnotesize inclusive}) }{dH_T \, dy^*}} 
\, .
\label{eq:rdphi}
\end{equation}
The denominator, $d^2\sigma_{\rm dijet}(\mbox{\small inclusive})/(dH_T\,dy^*)$,
is the cross section for the production of two or more jets,
with $p_T > \ptmin$, and $\yboost < \yboost^{\rm max}$,
double differentially in the variables $y^*$ and $H_T$.
The numerator, $d^2\sigma_{\rm dijet}(\Dphi < \Dphimax)/(dH_T \, dy^*)$,
is a subset of the denominator with the additional requirement
that the two leading $p_T$ jets have $\Dphi < \Dphimax$.
The quantity $\Rdphi$ is measured
as a function of the parameter $\Dphimax$, and in bins of
$\ystar$ and $H_T$, and therefore expressed as $\Rdphi(H_T, \ystar, \Dphimax)$.

It may be convenient to introduce an additional requirement 
of an $H_T$-dependent lower limit on the leading jet $p_T$
as $p_{T1} > f \cdot H_T$,
in both the numerator and the denominator. 
This requirement (which cuts the tail of low leading jet $p_T$)
is necessary in the experiment if events are triggered 
by inclusive single jet triggers.
The value of $f$ should not be too large, so as to not restrict
the multi-jet phase space too strongly.
We recommend to set $f$ not larger than $f = 1/3$, so that 
the phase space for $2\rightarrow 2$ and  $2\rightarrow 3$ processes
is not affected.

\subsection{Phase Space Scenarios for the LHC and the Tevatron
\label{sec:ps}}

To produce specific theory predictions, we propose two scenarios
of phase space regions
in which $\Rdphi$ can be measured at the LHC and the Tevatron.
While making realistic choices that take into account current 
practices by the experiments, we try to keep the two scenarios 
as similar as possible, so that the results can be used 
to study the $\sqrt{s}$ dependence of $\Rdphi$.

\paragraph{LHC Scenario}

We assume the running conditions of 2012, where the LHC was
producing $pp$ collisions at $\sqrt{s}= 8\,$TeV.
Following the choices by the ATLAS and CMS experiments,
jets are defined using the anti-k$_t$ jet algorithm~\cite{Cacciari:2008gp},
here with a jet radius of $R=0.6$ (in the $y$-$\phi$ plane),
which is within the range of 0.4--0.7 that is used by ATLAS and CMS.
The parameters in the $\Rdphi$ definition are set to
$\ystarmax = 2.0$, $\yboost^{\rm max} = 0.5$, 
and $\ptmin = 100\,$GeV.
The additional $H_T$-dependent requirement on the leading
jet $p_T$ is $p_{T1}/H_T > 1/3$.
The $\ystarmax$ and $\yboost^{\rm max}$ requirements ensure
that the two leading $p_T$ jets and all other jets entering
the $H_T$ sum are well-contained in the detector, within $|y|<2.5$.
The $\ptmin$ requirement ensures that all jets are well measured
in the experiment, and that pileup contributions and
non-perturbative corrections are small.
For studies of the $\sqrt{s}$ dependence of $\Rdphi$, we note that 
the $\ptmin$ requirement translates to a requirement for the
scaling variable $x_T = 2p_T/\sqrt{s}$ of $x_{T \rm min} = 0.025$.
The parameter $\Dphimax$ is set to $7\pi/8$, $5\pi/6$, or $3\pi/4$,
and the $y^*$ regions are chosen as $0<\ystar<0.5$, $0.5<\ystar<1$,
and $1<\ystar<2$.
The $H_T$ dependence is studied over the range $750 < H_T < 4000\,$GeV.

\paragraph{Tevatron Scenario}

In Run II, the Tevatron collided protons and anti-protons
at $\sqrt{s}= 1.96\,$TeV.
For the majority of the jet results, the CDF and D\O\ experiments 
use iterative seed-based cone algorithms with a cone of 
radius $\Rcone =0.7$ in $y$ and $\phi$.
For these studies, we apply the Run~II midpoint cone jet 
algorithm~\cite{run2cone} that is used by D\O.
We use the same values for the parameters 
$\ystarmax$, $\yboost^{\rm max}$, and $\Dphimax$,
the identical $y^*$ regions, and the same $p_{T1}/H_T$ requirement
as in the LHC scenario.
The only differences are the 
value of the $\ptmin$ requirement, which is set to $\ptmin = 30\,$GeV,
and the $H_T$ range of $180$--$900$\,GeV.
The $\ptmin$ requirement translates to a requirement for the
scaling variable of $x_{T \rm min} \approx 0.0306$,
which is slightly higher than the corresponding requirement
in the LHC scenario (of $x_{T \rm min} = 0.025$).
However, we use this value because
it corresponds to the
lower $p_T$ requirements used in recent multi-jet measurements 
at the Tevatron~\cite{:2012xi,:2012sy}

\begin{table}[tbp]
\centering
\begin{tabular}{|c|c|c|}
\hline
 & LHC  & Tevatron \\
 & $pp$ at $\sqrt{s}= 8\,$TeV & $\ppbar$ at $\sqrt{s}= 1.96\,$TeV \\
\hline 
jet algorithm & anti-k$_t$, $R=0.6$    & Run II cone, $\Rcone=0.7$ \\
$\ptmin$ ($x_{T \rm min}$) & 100\,GeV  (0.0250) & 30\,GeV ($\approx$ 0.0306) \\
$\yboost^{\rm max}$ & 0.5 & 0.5 \\
$\ystarmax$  & 2.0 & 2.0 \\
$p_{T1}/H_T$ & $>1/3$ & $>1/3$ \\
$\Dphimax$  & $7\pi/8$, $5\pi/6$, $3\pi/4$ & $7\pi/8$, $5\pi/6$, $3\pi/4$ \\
$\ystar$ ranges & 0.0--0.5, 0.5--1.0, 1.0--2.0 & 0.0--0.5, 0.5--1.0, 1.0--2.0\\
$H_T$ range & 750--4000\,GeV & 180--900\,GeV \\
\hline
\end{tabular}
\caption{\label{tab:ps} Summary of the phase space definitions
for the LHC and the Tevatron scenarios.}
\end{table}

The parameters, defining the phase space for the LHC and the Tevatron
scenarios are summarized in table~\ref{tab:ps}.

\section{Theory Predictions and their Properties}
\label{sec:thy}

In this section, we compute the perturbative and the non-perturbative 
contributions for $\Rdphi$ and investigate their properties.
We compare the predictions for the LHC and the Tevatron scenarios,
and investigate the differences due to the differences in $\sqrt{s}$,
and due to the slightly different phase space requirements.

All theory results have been obtained using the implementations
of the anti-k$_t$ and the D\O\ Run II cone jet algorithms in
{\sc fastjet}~\cite{Cacciari:2011ma,Cacciari:2005hq}.

\subsection{NLO pQCD Predictions \label{sec:nlo}}

\begin{figure}  
\centering
\includegraphics[scale=1]{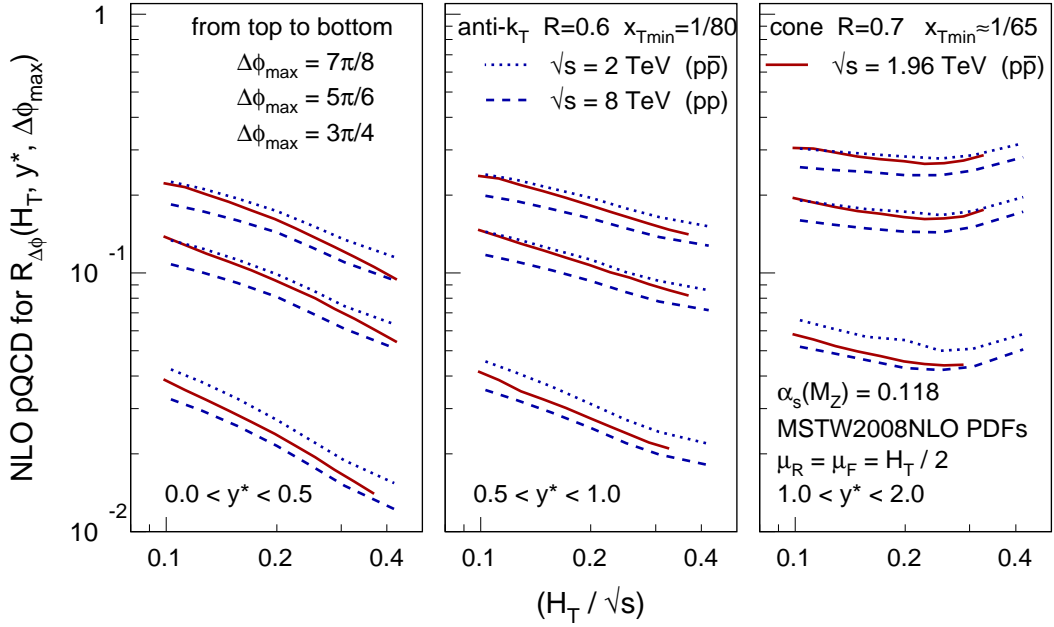}
\caption{\label{fig:sqrts} The NLO pQCD predictions for $\Rdphi$
as a function of $(H_T/\sqrt{s})$, and in regions of $\ystar$ 
(columns) and for different $\Dphimax$ requirements.
The results are shown for the LHC and the Tevatron scenarios,
and for a LHC-like scenario in which the LHC scenario
is scaled to $\sqrt{s}=2\,$TeV and modified to $\ppbar$ collisions.
}
\end{figure}

The NLO (LO) pQCD predictions for $\Rdphi$ are computed by taking
the ratios of the NLO (LO) pQCD predictions for the cross sections
in the numerator and the denominator in eq.~\eqref{eq:rdphi}.
The denominator is the inclusive dijet cross section
for which the NLO (LO) prediction is computed at 
$\ord(\as^3)$ ($\ord(\as^2)$).
Due to the additional requirement of $\Dphi < \Dphimax$, 
the numerator receives only contributions from final states
with three or more jets.
Therefore, the numerator is a three-jet cross section
for which the NLO (LO) prediction is computed at 
$\ord(\as^4)$ ($\ord(\as^3)$).
All NLO and LO pQCD results are computed using 
{\sc nlojet++}~\cite{Nagy:2003tz,Nagy:2001fj}, 
interfaced to {\sc fastnlo}~\cite{Kluge:2006xs}.
The calculations are made in the $\overline{\mbox{MS}}$ 
scheme~\cite{Bardeen:1978yd} for five active quark flavors,
and using the next-to-leading logarithmic (two-loop) approximation 
of the renormalization group equation.
The value of $\as(M_Z)=0.118$ is used consistently in the 
matrix elements and in the MSTW2008NLO PDF sets~\cite{Martin:2009iq}.
The central choice $\mu_0$ for the renormalization and 
factorization scales is $\mur = \muf = \mu_0 = H_T/2$.\footnote{At LO, 
this choice coincides with the common choices of $\mu_{R,F} = p_T$ for
inclusive jet production and $\mu_{R,F} = (p_{T1} +p_{T2})/2$ for
dijet production.}

The results of the NLO calculations for the LHC and the Tevatron scenarios
are displayed in 
figure~\ref{fig:sqrts}, where $\Rdphi$ is shown as a function
of $(H_T / \sqrt{s})$ in different regions of $\ystar$ and for different
$\Dphimax$ requirements.
In different regions of $H_T$ and $\ystar$, and for 
different choices of $\Dphimax$, 
$\Rdphi$ has values in the range 0.012--0.32.
In most phase space regions, $\Rdphi$ decreases with increasing $H_T$,
except at $1< \ystar < 2$ where $\Rdphi$ increases
again at high $H_T$.
At fixed $H_T$, $\Rdphi$ increases with increasing $\ystar$.
The fact that $\Rdphi$ decreases with decreasing $\Dphimax$ is a trivial
phase space effect, since a stronger $\Dphi$ requirement 
leads to a smaller cross section in the numerator.

For a fixed $x_{T \rm min}$ requirement and at fixed $(H_T / \sqrt{s})$,
the $\sqrt{s}$ dependence of the perturbative results for $\Rdphi$
is only introduced
through the evolution of $\as$ and the PDFs with the scales 
$\mur$ and $\muf$.
In figure~\ref{fig:sqrts},
the $\sqrt{s}$ dependence of $\Rdphi$ cannot directly be judged 
based on the comparison of the LHC and the Tevatron scenarios,
as the two differ in the $x_{T \rm min}$ requirement and in the jet algorithm.
The following study is made to separate the latter effects 
from the genuine $\sqrt{s}$ dependence of $\Rdphi$.
Using the flexibility provided by {\sc fastnlo}, we use the 
{\sc fastnlo} coefficient tables
for the LHC scenario (for $pp$ collisions at $\sqrt{s}=8\,$TeV)
to compute the corresponding predictions for $\ppbar$ collisions at 
the same $\sqrt{s}$.
The results for the latter (not shown in figure~\ref{fig:sqrts})
agree with those for the LHC scenario
better than 0.8\% for $H_T < 2\,$TeV
and always better than 3.2\% in the phase space studied,
meaning that $\Rdphi$ is insensitive to the difference
between $pp$ and $\ppbar$ initial states.
Then we use the {\sc fastnlo} results for the LHC scenario
to compute the corresponding predictions for a LHC-like scenario
(i.e.\ using the same jet algorithm and the same $x_{T \rm min}$ 
requirement)
for $\ppbar$ collisions at $\sqrt{s}=2\,$TeV.
These predictions are shown in figure~\ref{fig:sqrts}
as the dotted line.
The $\Rdphi$ results at $\sqrt{s}=2$\,TeV are 
10--20\% higher than those at 8\,TeV.
This $\sqrt{s}$ dependence is consistent with the running of $\as$ over a factor of four
in energy.\footnote{The PDFs approximately cancel in the ratio $\Rdphi$, 
so the $\muf$ dependence of the PDFs does not have a significant impact
on the $\sqrt{s}$ dependence of $\Rdphi$.}
For $\Dphimax = 7\pi/8$ and $5\pi/6$,
the results for the LHC-like scenario agree within 5\%
with those for the Tevatron scenario.
Only for $\Dphimax = 3\pi/4$  the differences become
larger (slightly more than 10\% at $\ystar > 1$).
From this we conclude that, even with different jet algorithms
and slightly different $x_{T \rm min}$ requirements,
a comparison of Tevatron and LHC data is probing the 
$\sqrt{s}$ dependence of $\Rdphi$ and testing the
corresponding theory predictions.

\begin{figure}  
\centering
\includegraphics[width=.85\textwidth]{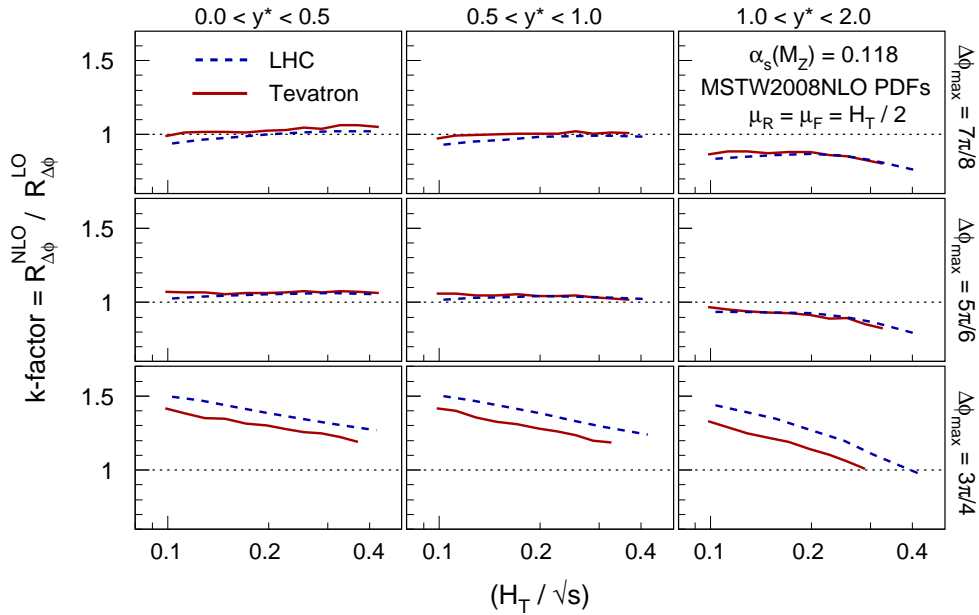}
\caption{\label{fig:kfac} 
The NLO $k$-factors for $\Rdphi$ as function of  $(H_T/\sqrt{s})$,
in different regions of $y^*$ (columns) and for different 
$\Dphimax$ (rows),
for the LHC and the Tevatron scenarios.
}
\end{figure}

In the following, we investigate the NLO $k$-factors and the scale dependence
as indicators for the stability of the perturbative expansion,
and we study the PDF uncertainties for $\Rdphi$.
The NLO $k$-factors are computed as the ratio of the NLO and the LO
predictions, $k = \Rdphi^{\rm NLO}/\Rdphi^{\rm LO}$.
The values of the $k$-factors are displayed in figure~\ref{fig:kfac}
as a function of $(H_T / \sqrt{s})$, for the LHC and the Tevatron scenarios.
For $\Dphimax = 7\pi/8$ and $5\pi/6$, the $k$-factors
for the LHC and the Tevatron are always close to unity;
they decrease slightly with increasing $\ystar$
and are almost independent of $H_T$.
Due to kinematic constraints, the region of $\Dphi < 2\pi/3$ 
is only accessible in four-jet final states.
For this reason, the kinematic region of $\Dphimax = 3\pi/4$ 
also receives large contributions from four-jet production 
which are only modeled at LO by the $\ord(\as^4)$ calculation 
for the numerator of $\Rdphi$.
This is reflected in the large NLO $k$-factors
for $\Dphimax = 3\pi/4$ which are as large as $k=1.5$ at lower $H_T$.

\begin{figure}  
\centering
\includegraphics[width=.8\textwidth]{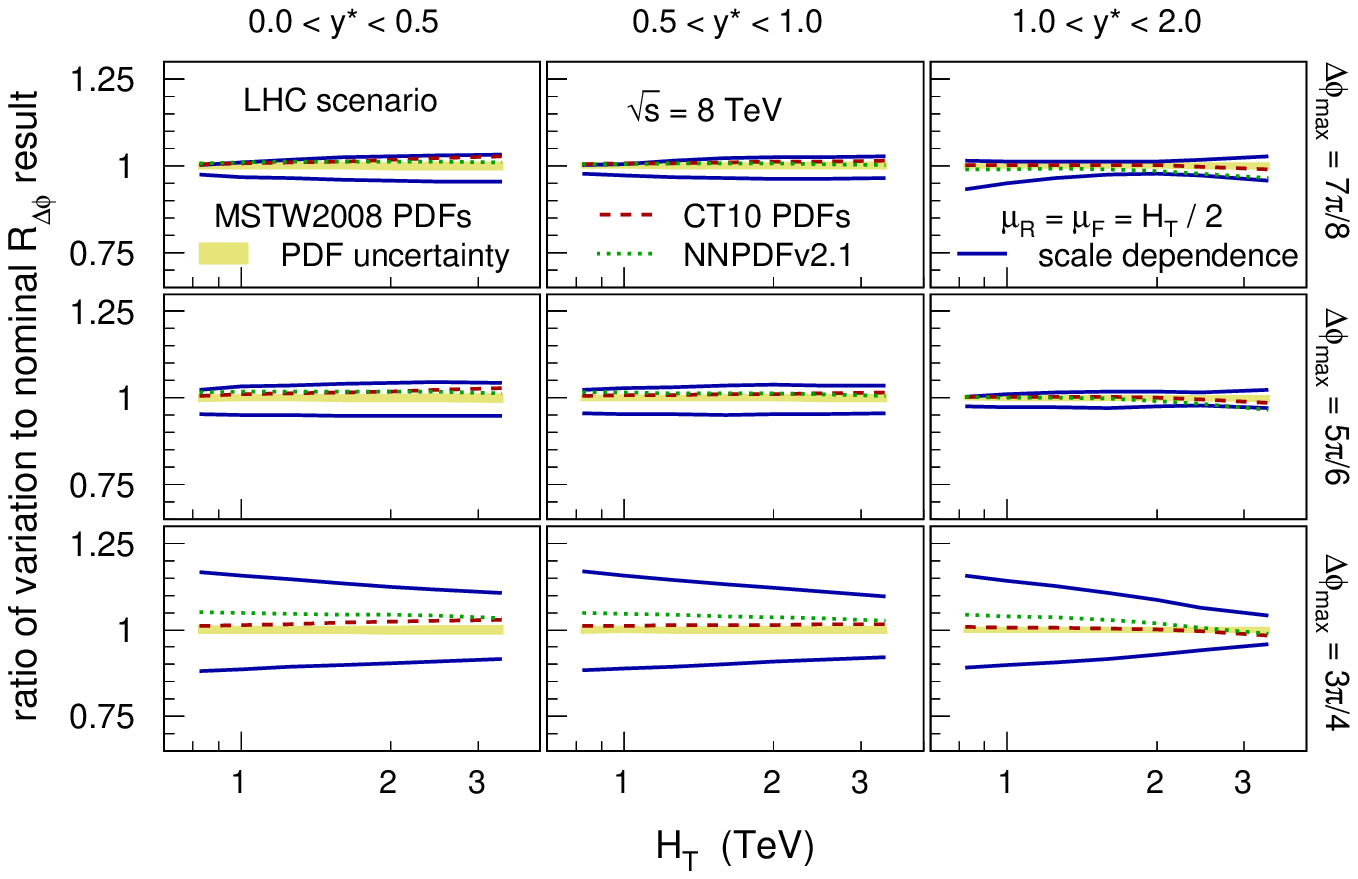}
\includegraphics[width=.8\textwidth]{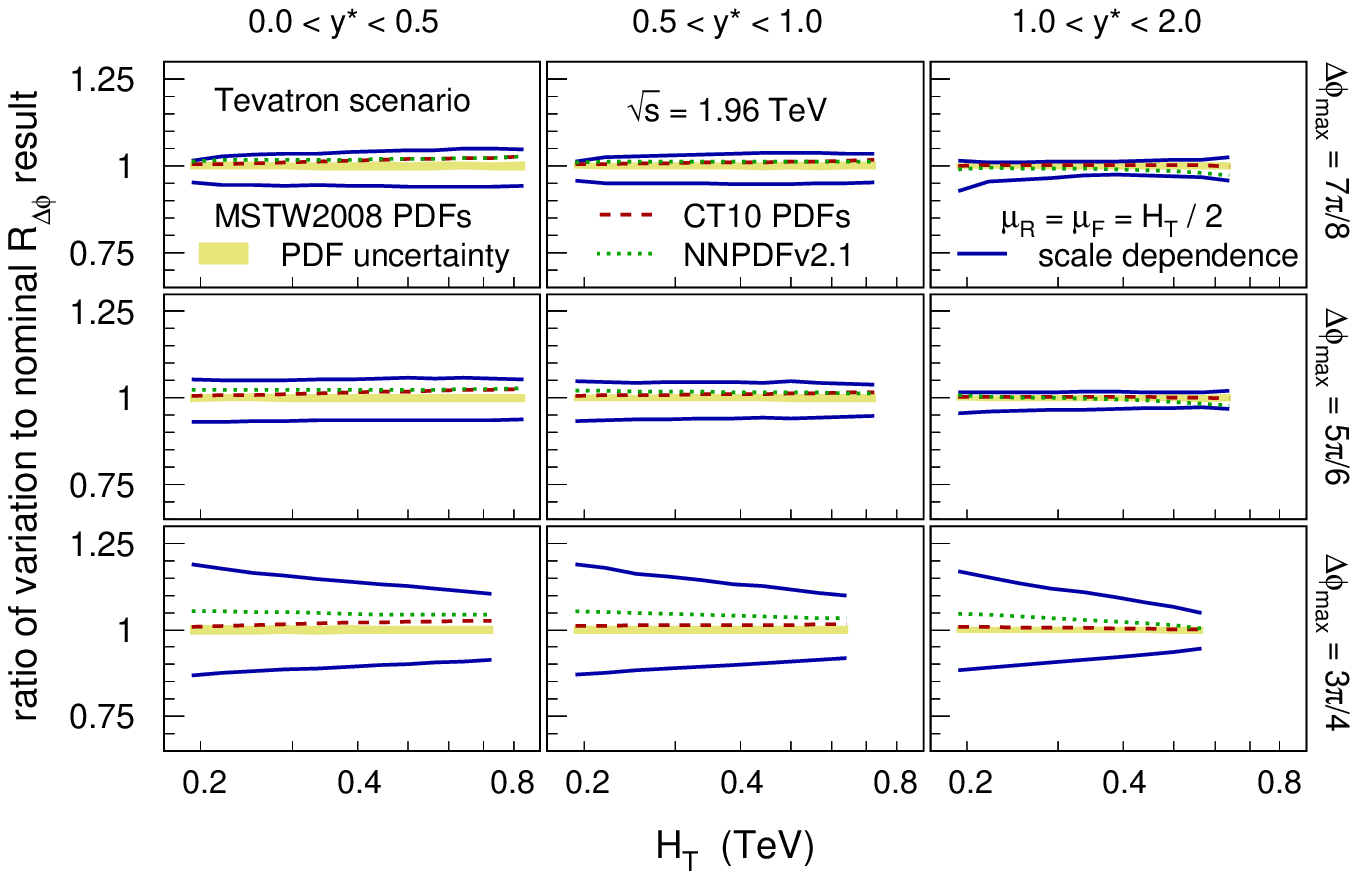}
\caption{\label{fig:thyunclhc} 
The renormalization and factorization scale dependence and the 
PDF uncertainties for the MSTW2008NLO PDFs
for the NLO pQCD predictions for the LHC and Tevatron scenarios.
}
\end{figure}

The uncertainties due to the scale dependence are computed
from the relative variations of the $\Rdphi$ results
when $\mur$ and $\muf$ are varied independently 
around $\mu_0 = H_T/2$
between $\mu_0 / 2 $ and $2\mu_0$ but never exceeding
$0.5 \le \mur / \muf \le 2.0$.
These uncertainties are displayed in figure~\ref{fig:thyunclhc}
for the LHC and Tevatron scenarios.
For the LHC (Tevatron) scenario, these uncertainties
are typically 5\% (7\%) for $\Dphimax = 7\pi/8$ and $5\pi/6$,
and up to 19\% (17\%) for $\Dphimax = 3\pi/4$.
The latter is directly related to the large NLO $k$-factors in this region.
The uncertainties for $\Dphimax = 7\pi/8$ and $5\pi/6$ are slightly smaller 
as compared to the uncertainties for other ratios of three-jet and 
dijet cross sections like $\Rtt$ and $\Rdr$ as recently measured 
at the LHC~\cite{Aad:2011tqa} and the Tevatron~\cite{:2012sy,:2012xi}.

The PDF uncertainties are computed for the MSTW2008NLO PDFs using 
the up and down variations of the 20 orthogonal PDF uncertainty eigenvectors,
corresponding to the 68\% C.L.
The MSTW2008NLO PDF uncertainties for the LHC and Tevatron scenarios are shown 
in figure~\ref{fig:thyunclhc} and they are always below 1\%.
Also displayed in figure~\ref{fig:thyunclhc} are the NLO pQCD predictions 
obtained for CT10~\cite{Lai:2010vv} and NNPDFv2.1~\cite{Ball:2011mu} PDFs.
Those results are larger by up to 5\%
as compared to the results for MSTW2008NLO PDFs.
The largest deviations occur at smallest $\Dphimax$.

From these studies we conclude that theory predictions are most reliable
(as indicated by a small scale dependence and $k$-factors which 
are close to unity) in the kinematic regions of $\ystar < 1$
and for $\Dphimax =7\pi/8$ and $5\pi/6$.

\subsection{Non-perturbative Effects}

\begin{figure}  
\centering
\includegraphics[width=.84\textwidth]{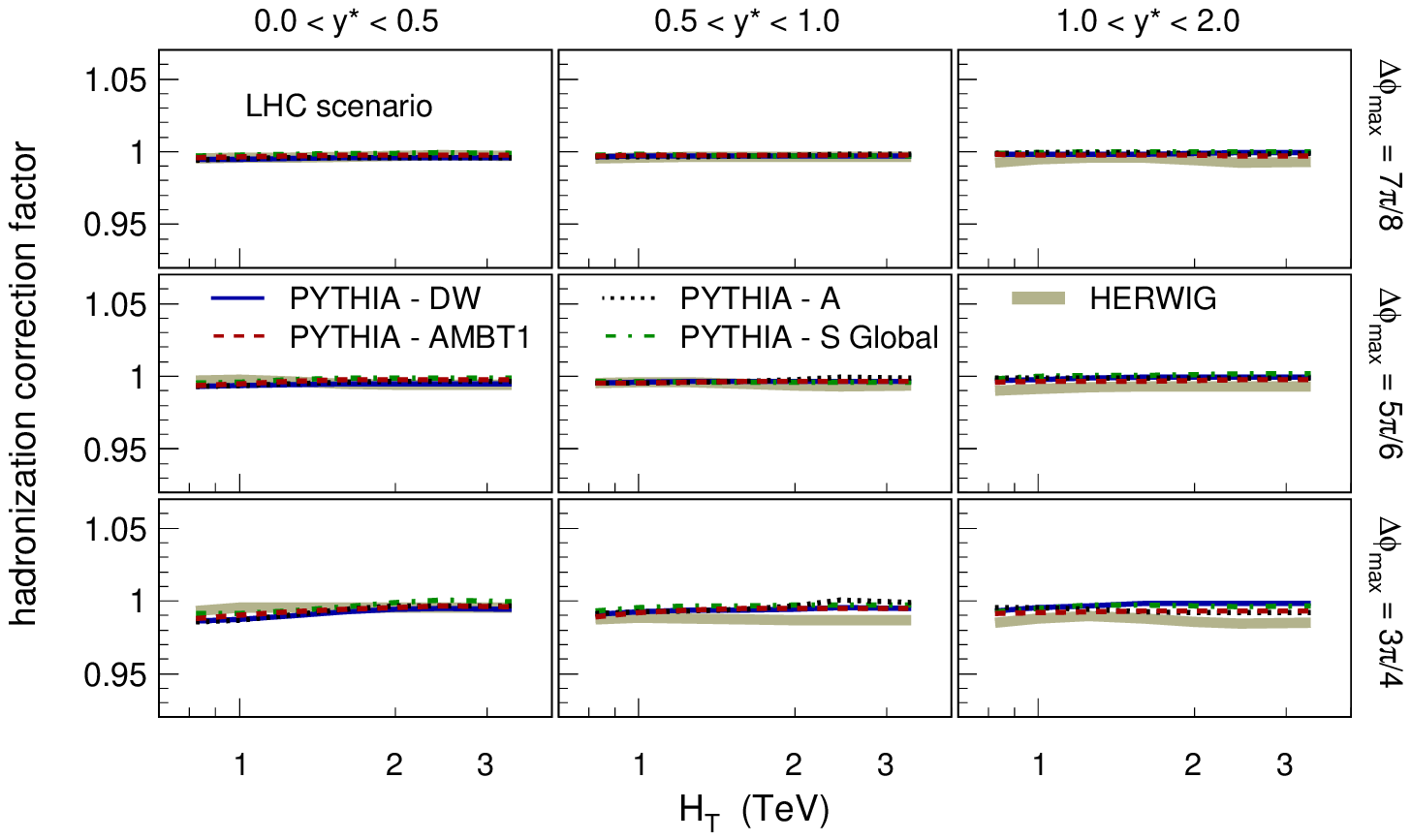}
\includegraphics[width=.84\textwidth]{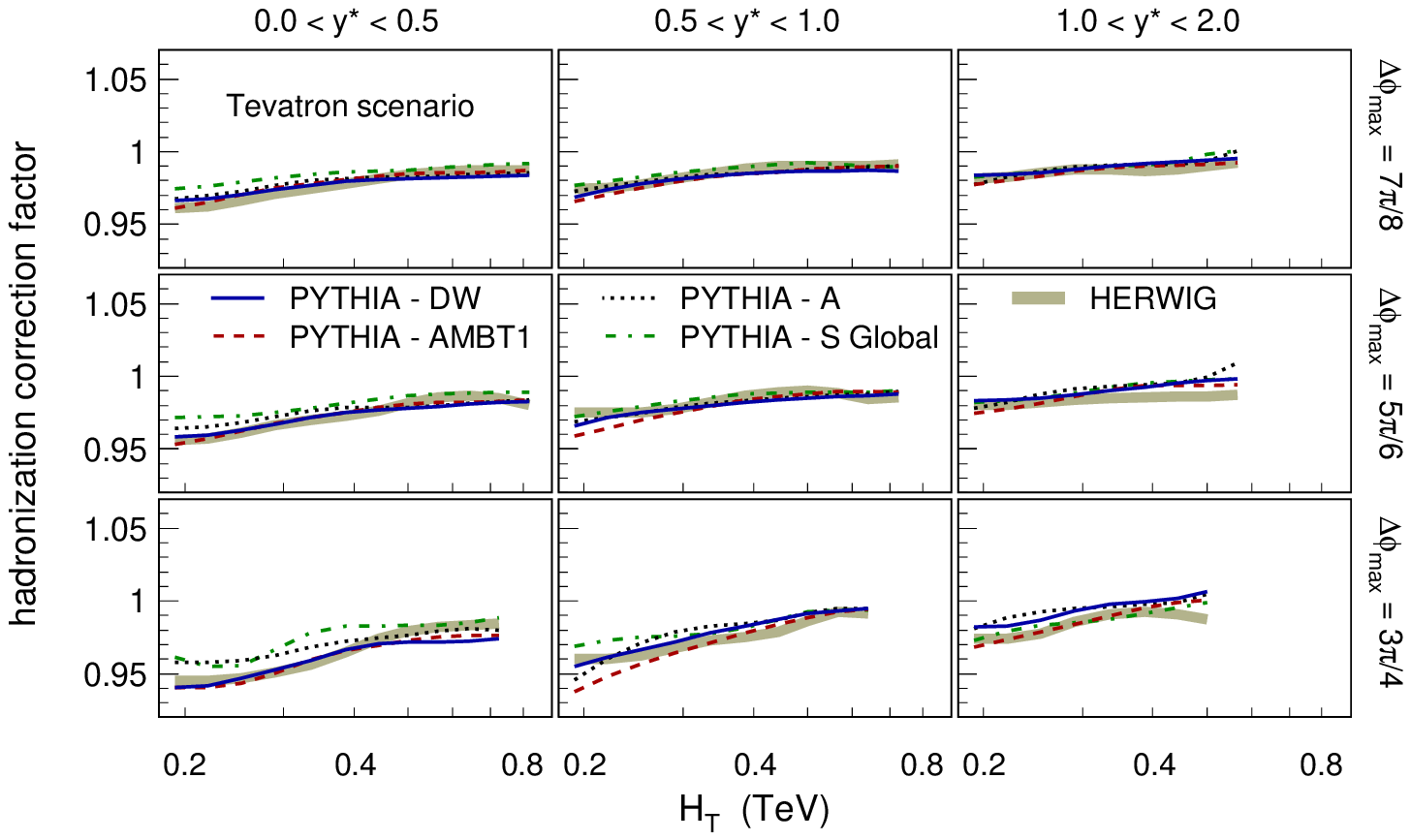}
\caption{\label{fig:hadlhc} The hadronization corrections 
  for $\Rdphi$, plotted as a function of $H_T$
  in different $y^*$ regions (columns) and for different
  values of $\Dphimax$ (rows), for the LHC and Tevatron scenarios.}
\end{figure}

\begin{figure}  
\centering
\includegraphics[width=.84\textwidth]{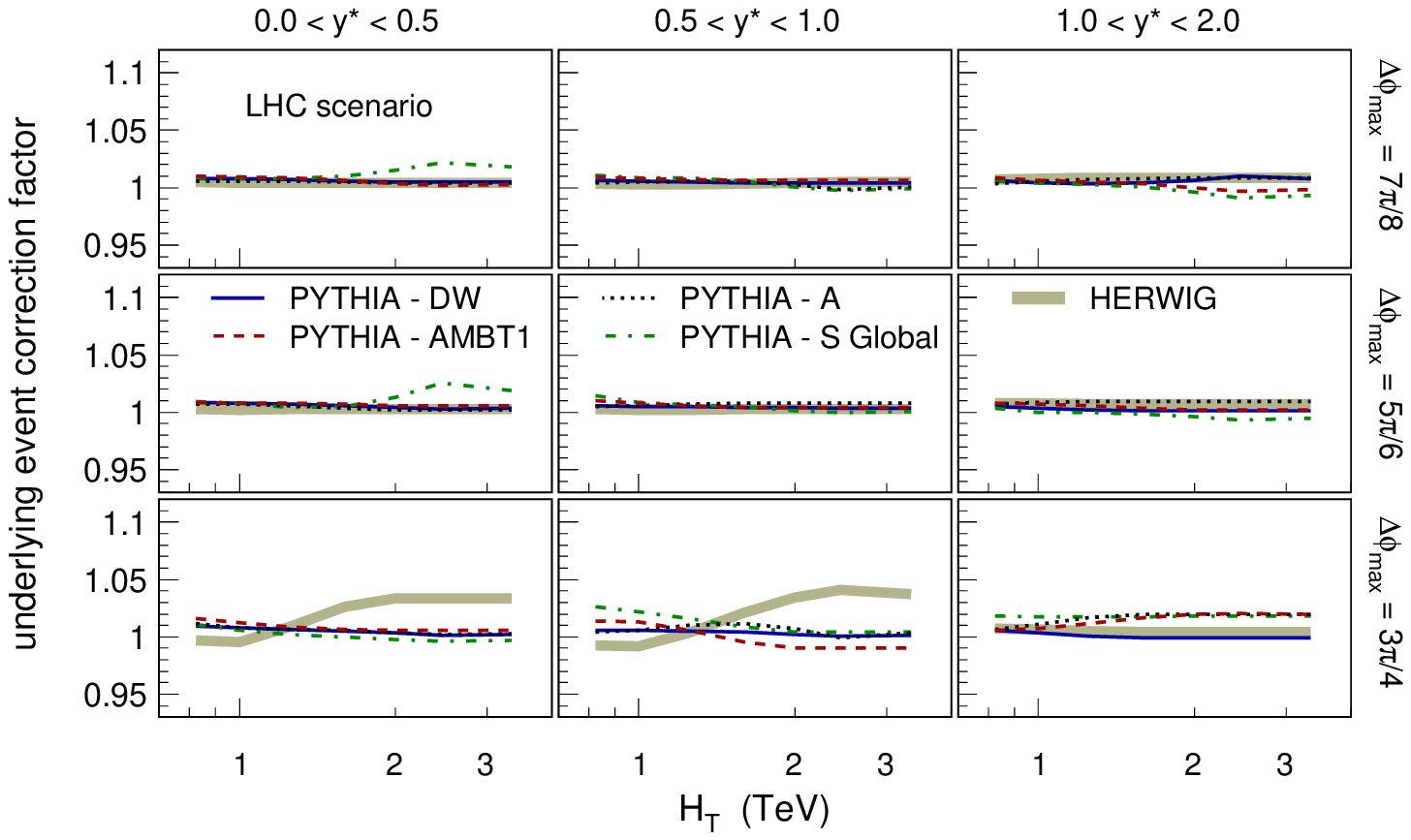}
\includegraphics[width=.84\textwidth]{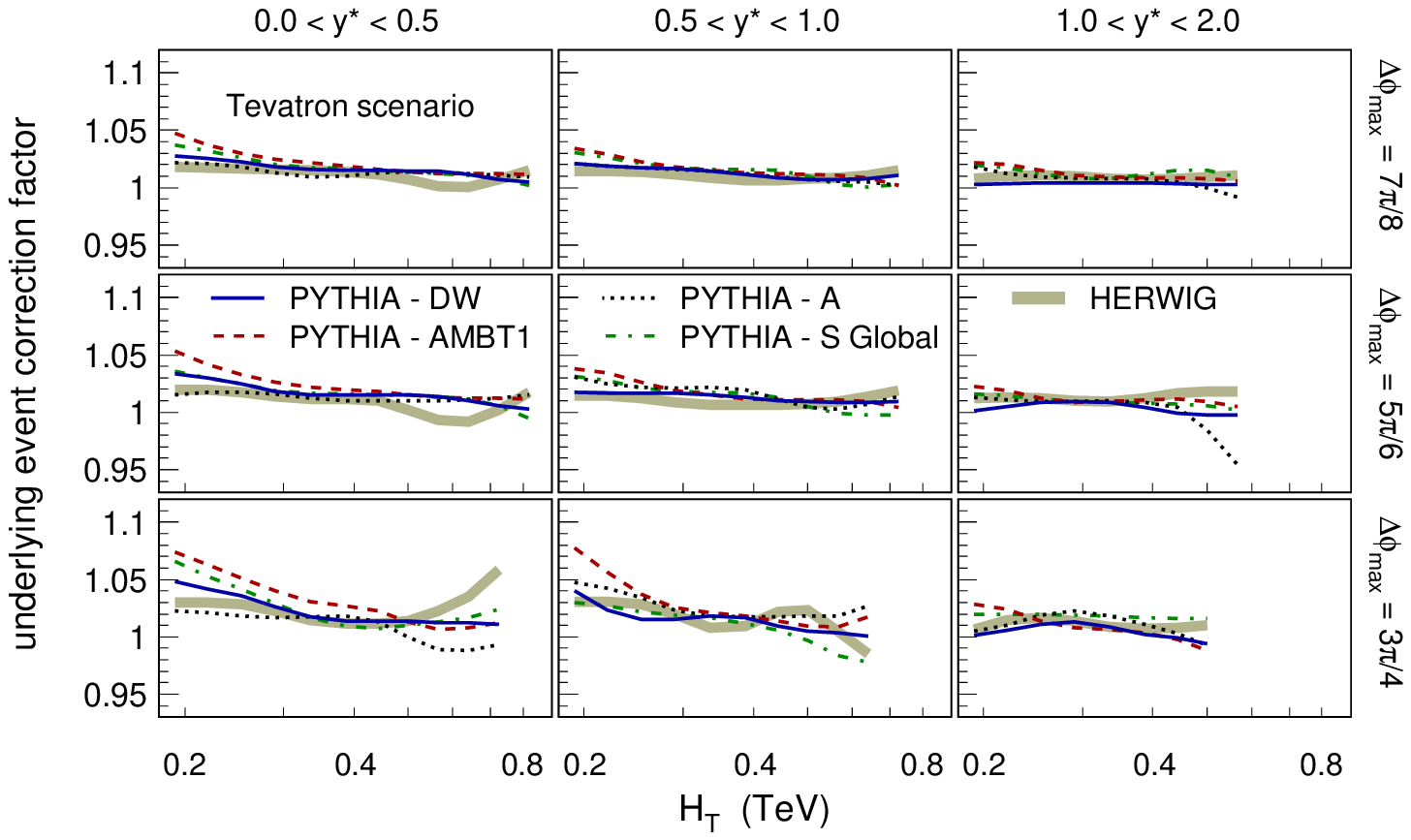}
\caption{\label{fig:uelhc}  The underlying event corrections 
  for $\Rdphi$, plotted as a function of $H_T$
  in different $y^*$ regions (columns) and for different
  values of $\Dphimax$ (rows), for the LHC and Tevatron scenarios.}
\end{figure}

In these studies, we consider non-perturbative effects
due to the underlying event and due to hadronization corrections.
Both corrections are estimated using the models implemented in the 
event generators \herwig\/~6.520~\cite{Corcella:2000bw,Corcella:2002jc}
and \pythia\/~6.426~\cite{pythia,Sjostrand:2006za}.
The \herwig\ results are obtained using default settings
and the \pythia\ results are obtained for four different popular tunes.
These are tunes DW~\cite{Albrow:2006rt} and A~\cite{tuneA},
which use a $Q^2$ ordered parton shower and an older model 
for the underlying event, and the tunes 
AMBT1~\cite{Diehl:2010zz} and S~Global~\cite{Schulz:2011qy},
which use a $p_T$ ordered parton shower and the new model for
the underlying event~\cite{Sjostrand:2004pf,Sjostrand:2004ef}.
The non-perturbative corrections are obtained from the results 
of three calculations in which $\Rdphi$ is computed
\begin{itemize}
\item[(1)] at the parton level (the partons after the parton shower) 
           with no underlying event,
\item[(2)] at the particle level (using all stable particles) 
           with no underlying event, and
\item[(3)] at the particle level (using all stable particles) 
           with underlying event.
\end{itemize}
The total non-perturbative correction  $c_{\rm npert}$
is defined as the product of
the hadronization correction $c_{\rm hadr}$ and the underlying event 
correction $c_{\rm ue}$ 
which are each given by ratios of $\Rdphi$ results 
on different levels, as
\begin{equation}
c_{\rm npert} = c_{\rm hadr} \cdot c_{\rm ue} 
\hskip7mm \mbox{with} \hskip7mm
c_{\rm hadr} = \frac{\Rdphi^{(2)}}{\Rdphi^{(1)}}
\hskip7mm \mbox{and} \hskip7mm
c_{\rm ue} = \frac{\Rdphi^{(3)}}{\Rdphi^{(2)}}   \, .
\end{equation}
Figure~\ref{fig:hadlhc} shows the hadronization corrections 
for $\Rdphi$ for the LHC and the Tevatron scenarios.
The hadronization corrections for the LHC are very small and always below 1.5\%
($0.985 < c_{\rm hadr} < 1.00$) at all $H_T$, $y^*$, and for
all $\Dphimax$ requirements.
The hadronization corrections for the Tevatron, although slightly larger,
are still always below 6\% ($0.94 < c_{\rm hadr} < 1.01$).
The \herwig\ results and the \pythia\ results for the different tunes 
agree always within 1\% (3\%), for the LHC (Tevatron) scenario.

The corrections for $\Rdphi$ due to effects from the underlying event 
are displayed in figure~\ref{fig:uelhc} for the LHC and the Tevatron scenarios.
For the LHC, the underlying event corrections are always below 4\%
($0.99 < c_{\rm ue} < 1.04$), at all $H_T$, $y^*$, and for all 
$\Dphimax$ requirements.
The maximum corrections increase with decreasing $\Dphimax$.
They are 2\% for $\Dphimax = 7\pi/8$, 3\% for $\Dphimax = 5\pi/6$,
and 4\% for $\Dphimax = 3\pi/4$.
For the Tevatron, the underlying event corrections  
($0.99 < c_{\rm ue} < 1.08$) have the same qualitative behavior 
and they are approximately twice as large as those for the LHC.
The \herwig\ results and the \pythia\ results for the different tunes 
are in good agreement.
The different model predictions for the LHC (Tevatron) scenario
agree better than 2\% (3\%) for $\Dphimax = 7\pi/8$,
and always better than 4\% (6\%).

The total non-perturbative corrections are always in the range 
$0.98 < c_{\rm npert} < 1.03$ ($0.96 < c_{\rm npert} < 1.03$) 
for the LHC (Tevatron) scenario.
The smallness of these corrections and their small model dependence
are remarkable features of the quantity $\Rdphi$ which makes $\Rdphi$ 
well-suited for precision tests of pQCD.

\section{Phenomenology \label{sec:pheno}}

In this section, we discuss two examples of the potential impact 
of future $\Rdphi$ measurements for QCD phenomenology.

\subsection[Sensitivity to $\as$]{\boldmath Sensitivity to $\as$}  

In pQCD, $\Rdphi$ is computed as a ratio of three-jet and dijet cross sections,
which is, at LO, proportional to $\as$.
In the following, we study the sensitivity of $\Rdphi$ to $\as$
and investigate the effects of theoretical and experimental uncertainties
on the $\as$ results.
These studies are made in the kinematic region of $\ystar < 0.5$ and
for $\Dphimax = 7\pi/8$ where the pQCD predictions are most
reliable (see section~\ref{sec:nlo}).

In all studies, $\as$ is varied consistently in the pQCD matrix elements
and in the PDF sets.
The continuous dependence of the NLO pQCD predictions for $\Rdphi$ 
on $\as$ is obtained using cubic interpolation between the 
discrete $\asmz$ values for which the MSTW2008NLO PDFs sets are 
available.\footnote{The MSTW2008NLO PDF sets are available for 
$\asmz = 0.110, \, 0.111, \, 0.112, \cdots, \, 0.130$.}
Where needed, $\asmz$ is converted from the scale $\mur = M_Z$ to 
the scale $\mur=H_T/2$, using the two-loop solution of the 
renormalization group equation.

\begin{figure}  
\centering
\includegraphics[scale=1]{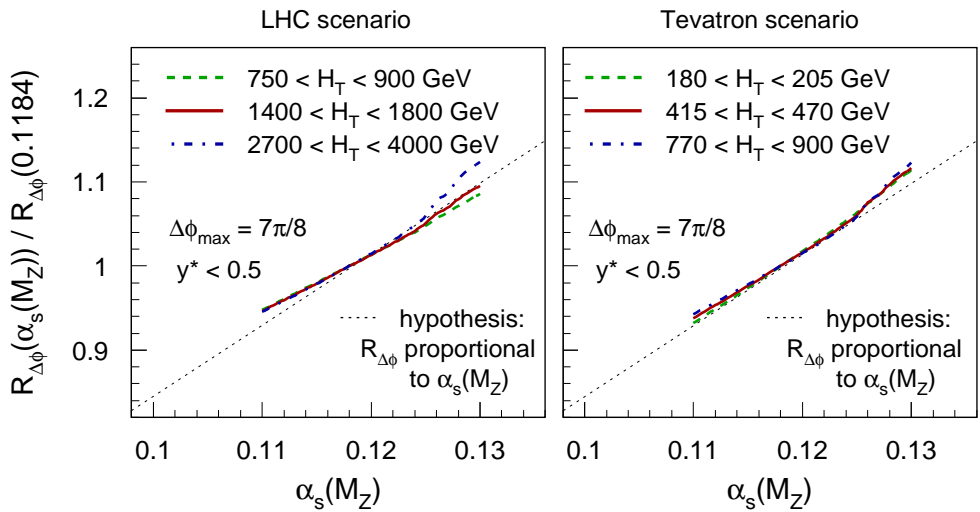}
\caption{\label{fig:asmzdep}  
The $\asmz$ dependence of $\Rdphi$,
normalized by the $\Rdphi$ value at $\asmz=0.1184$, for the LHC scenario (left) and the 
Tevatron scenario (right) in three different $H_T$ bins
for $\Dphimax =7\pi/8$ and $\ystar < 0.5$.}
\end{figure}

A first impression of the $\as$ sensitivity is obtained by studying 
the $\asmz$ dependence of the NLO pQCD predictions for $\Rdphi$.
For this purpose, we plot $\Rdphi$ for values of $\asmz = 0.110$--$0.130$
(labeled $\Rdphi(\asmz)$), normalized to the value of $\Rdphi$ 
for $\asmz = 0.1184$ (the world average value~\cite{PDG2012})
as a function of $\asmz$.
The results are shown in figure~\ref{fig:asmzdep} for three 
different $H_T$ bins, for the LHC and the Tevatron scenarios, 
and are compared to the naive expectation of a linear relation
($\Rdphi \propto \as$).
Deviations from a linear dependence could be due to three different effects.
\begin{itemize}
\item[1.] The naive expectation of a linear dependence stems from
     the LO picture, and is modified due to NLO corrections.
\item[2.] The naive expectation assumes a perfect cancellation of the PDFs,
      while residual PDF effects may lead to non-linearities.
\item[3.] While figure~\ref{fig:asmzdep} shows the $\asmz$ dependence,
     the calculations for $\Rdphi$ are made for the scale $\mur = H_T/2$,
     and the relation between  $\as(\mur)$ and $\asmz$ as a function of 
     $\asmz$ is not linear, and involving logarithms of $(\mur/M_Z)$.
\end{itemize}
For $\asmz \lesssim 0.125$, in the Tevatron scenario, the $\asmz$ dependence 
of $\Rdphi$ is almost linear for $180 < H_T < 205\,$GeV 
(i.e.\ where $\mur = H_T/2 \approx M_Z$).
The decrease of the slope (and therefore the increasing non-linearity) 
of the curves for higher $H_T$ is likely caused by the third effect.
The change of the slopes around $\asmz \approx 0.125$ is likely caused by 
the second effect.
The reduced slope towards high $H_T$ implies a slightly reduced 
sensitivity to $\asmz$ at the LHC.

\begin{figure}  
\centering
\includegraphics[scale=1]{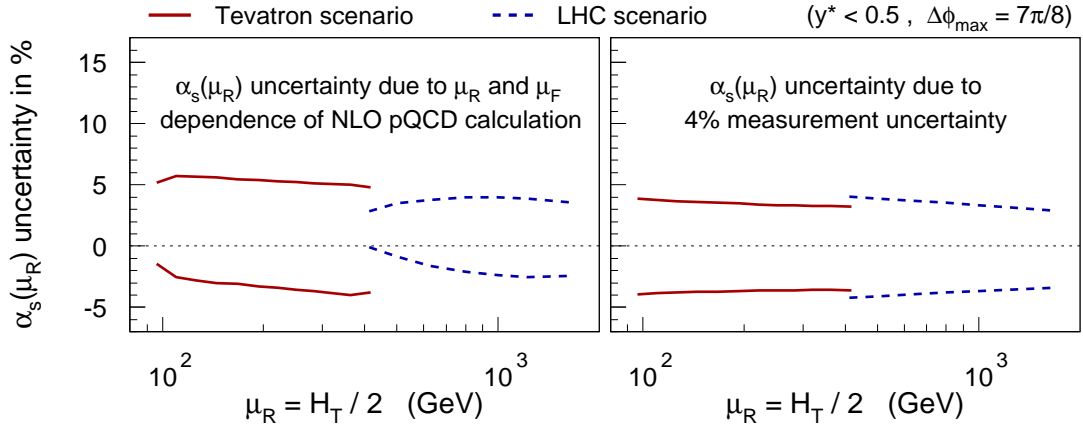}
\caption{\label{fig:assensit}  The estimated uncertainties
of $\as$ results extracted from $\Rdphi$ at the LHC and the Tevatron,
at a scale $\mur = H_T/2$, 
due to the renormalization and factorization scale dependence
of the NLO pQCD calculation (left) 
and due to an experimental uncertainty of 4\%,
for $\Dphimax =7\pi/8$ and $\ystar < 0.5$.}
\end{figure}

Currently, the precision of $\as$ results obtained from hadron colliders
which are based on NLO calculations, is limited by theory uncertainties 
stemming from the renormalization and factorization scale dependencies of the 
calculations.\footnote{The presently most precise $\as$ result from a hadron
collider was obtained using theory calculations beyond NLO
(adding the 2-loop corrections from threshold corrections) 
and has therefore smaller scale uncertainties~\cite{Abazov:2009nc}.
These contributions are, however, only available for inclusive jet production
and neither for dijet nor for three-jet production.}
Therefore we estimate the corresponding uncertainties for 
$\as$ extractions from $\Rdphi$.
In the typical procedure of most $\as$ analyses, 
the central $\as$ results are derived for a fixed
choice of the renormalization and factorization scales.
The uncertainties of $\as$ due to the scale dependence are then obtained
by repeating the $\as$ fits for variations of the scales around their
central values.
In the absence of actual $\Rdphi$ data, we estimate the corresponding 
uncertainties for $\as$ by computing the variations in $\as$ which are 
required to bring the NLO pQCD results at a different scale into agreement 
with those at the central scale.
As discussed in section~\ref{sec:nlo}, we use a central 
scale of $\mur = \muf = \mu_0 = H_T/2$ and a range of variations
in which $\mur$ and $\muf$ are varied independently
between $\mu_0/2$ and $2\mu_0$, while never exceeding
$0.5 \le \mur/\muf \le 2.0$.
The largest effects of all variations are quoted as the 
corresponding uncertainties for $\as$.
The expected uncertainties for $\as(\mur=H_T/2)$, derived using this procedure
are shown in figure~\ref{fig:assensit} (left)
in the kinematic range $\ystar < 0.5$ and $\Dphimax = 7\pi/8$,
as a function of $\mur$, for the LHC and the Tevatron.
For the LHC scenario, the uncertainties are between $-3\%$ and $+4\%$ 
at high $H_T$, and slightly lower at low $H_T$.
For the Tevatron, the uncertainties are between $-4\%$ and $+6\%$.
This is a good theoretical precision for testing the running of $\as$ 
at highest energies.

For comparison, we have also computed the uncertainty of $\as(H_T/2)$
resulting from an experimental uncertainty of 4\%.
The results in figure~\ref{fig:assensit} (right) show
that for the LHC and the Tevatron scenarios
this uncertainty is approximately of the same size
as the uncertainty due to the scale dependence.
In other words, if the total uncertainty for the $\as$ results should 
not be limited by the experimental precision, the experiments
must measure $\Rdphi$ with a precision of at least approximately 4\%.
Given the precision of recently published measurements of the multi-jet 
cross section ratios $\Rtt$~\cite{Chatrchyan:2011wn,Aad:2011tqa,:2012sy} 
and $\Rdr$~\cite{:2012xi}, this should be achievable.

\subsection{Event Generator Tuning}

\begin{figure}  
\centering
\includegraphics[scale=1]{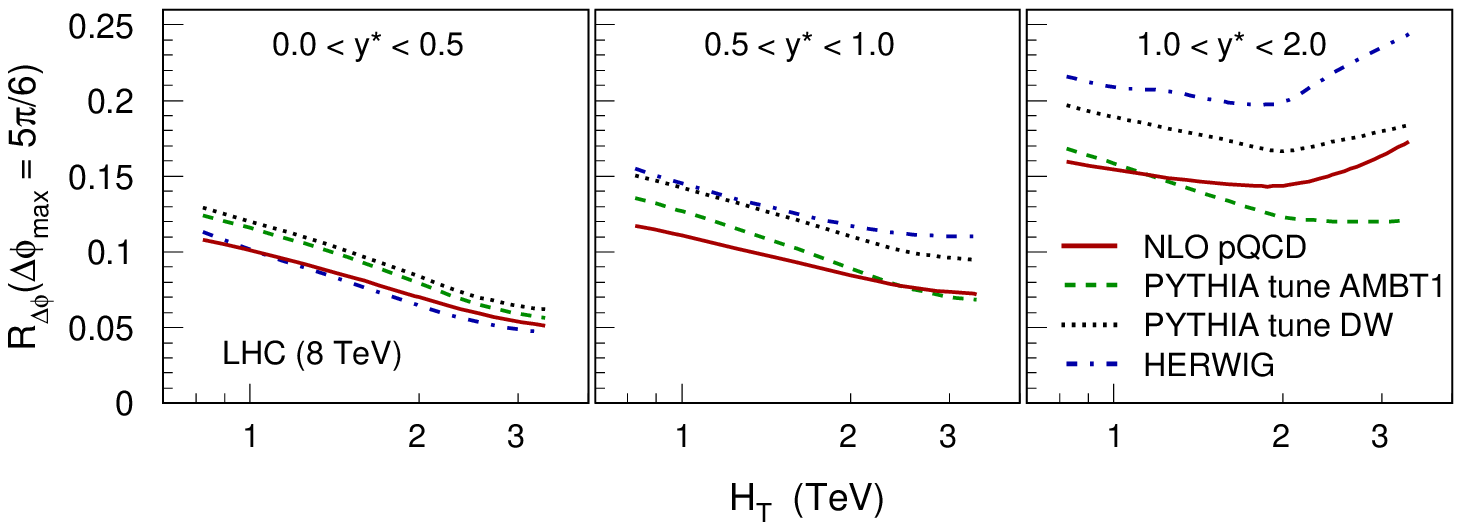}
\includegraphics[scale=1]{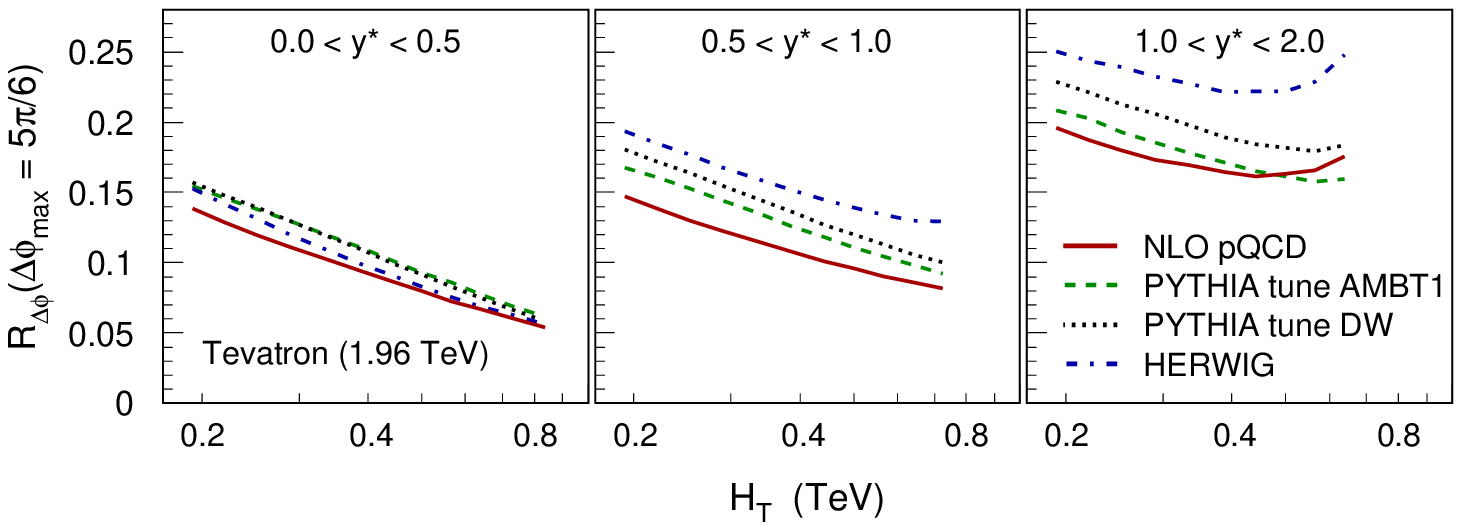}
\caption{\label{fig:mclhc}  
  Comparison of $\Rdphi$ predictions for $\Dphimax = 5\pi/6$
  of different event generators with the NLO pQCD predictions 
  as a function of $H_T$ in different $y^*$ regions
  for the LHC and Tevatron scenarios.}
\end{figure}

We also compute $\Rdphi$ predictions using the
Monte Carlo event generators \herwig\ 6.520 and \pythia\ 6.426.
The \herwig\ generator is used with default settings,
and for \pythia\ we use tune AMBT1 (derived by the ATLAS collaboration
using LHC data) and tune DW (which has been tuned to the previous 
D\O\ measurement of dijet azimuthal decorrelations).
The results for $\Dphimax = 5\pi/6$ are shown in figure~\ref{fig:mclhc}
for the LHC and the Tevatron scenarios.
At central rapidities (i.e.\ low $\ystar$) the predictions of the different
generators agree reasonably well with each other, and also with the 
NLO pQCD results.
The different generators (and tunes), however, predict very different $\ystar$ dependencies.
In the forward region ($1< \ystar < 2$), the predictions
differ strongly in magnitude and in shape.
The \herwig\ predictions are more than 30\% higher,
and the \pythia\ tune AMBT1 predictions have a very different shape
as compared to the NLO pQCD results.
The large range of the predictions from different \pythia\ tunes
and their differences to \herwig\ indicates that 
measurements of $\Rdphi$ at the LHC and at the Tevatron
will have strong impact on the future tuning of event generators.

\section{Summary}  

We have introduced a new quantity $\Rdphi$ for studies of the rapidity and 
transverse momentum dependence of dijet azimuthal decorrelations 
at hadron colliders.
Since $\Rdphi$ is defined as a ratio of cross sections, 
experimental and theoretical uncertainties,
which are correlated between the numerator and the denominator,
will cancel to a large extent.
We suggest to measure the rapidity and transverse momentum dependence 
of $\Rdphi$ using the longitudinally boost invariant variables $\ystar$
and $H_T$.
In pQCD, $\Rdphi$ is given by a ratio of three-jet and dijet cross sections,
and, at leading order, proportional to $\as$.
We have proposed scenarios for measuring $\Rdphi$ at the LHC and at the 
Tevatron, 
for which we have computed the NLO pQCD predictions and the
size of non-perturbative corrections.
The non-perturbative corrections are below 3\%
with a model dependence of typically less than 2\%.
The NLO pQCD predictions have PDF uncertainties of less than 1\%,
and a scale dependence of 4--6\% (for $\Dphimax = 7\pi/8$ and $5\pi/6$).
These properties make $\Rdphi$, and its $H_T$ dependence 
an ideal quantity for determinations of $\as$ and for
studying its running up to the energy frontier.

While these studies have focused on the $H_T$ dependence
of $\Rdphi$, we have also investigated the theoretical predictions
for the rapidity dependence.
We have shown that, at fixed $H_T$, 
NLO pQCD and the event generators \pythia\ and \herwig\
all predict an increase of $\Rdphi$ with $\ystar$, however,
the rate of the increase differs strongly between the different models.
Measurements of $\Rdphi$ at the LHC an the Tevatron will be able
to distinguish between the different predictions, test NLO pQCD and
play an important role in future tuning of Monte Carlo event generators.

\acknowledgments
We are thankful to Don Lincoln for numerous discussions and for 
comments on the manuscript.
This work has been supported by grants 
DE-FG02-99ER41117 and DE-FG02-10ER46723
from the U.S. Department of Energy.



\begin{thebibliography}{99}

\bibitem{PDG2012}
J.~Beringer et al. (Particle Data Group), 
    \emph{The Review of Particle Physics,}
    \emph{Phys.\ Rev.\ D} {\bf 86} (2012) 010001. 

\bibitem{Abazov:2011ub} 
  D0 Collaboration, V.~M.~Abazov { et al.},
  \emph{Measurement of three-jet differential cross sections 
     $d\sigma_{\text{3jet}} / dM_{\text{3jet}}$ in $p\bar{p}$ collisions 
     at $\sqrt{s}=1.96$ TeV,}
  \emph{Phys.\ Lett.\ B} {\bf 704} (2011) 434
  [arXiv:1104.1986].

\bibitem{Chatrchyan:2011wn} 
   CMS Collaboration, S.~Chatrchyan { et al.},
  \emph{Measurement of the Ratio of the 3-jet to 2-jet Cross Sections in $pp$ Collisions at $\sqrt{s} = 7$ TeV,}
  \emph{Phys.\ Lett.\ B} {\bf 702} (2011) 336
  [arXiv:1106.0647].


\bibitem{Aad:2011tqa} 
  ATLAS Collaboration, G.~Aad { et al.},
  \emph{Measurement of multi-jet cross sections in proton-proton collisions at a 7 TeV center-of-mass energy,}
  \emph{Eur.\ Phys.\ J.\ C} {\bf 71}, 1763 (2011)
  [arXiv:1107.2092].

\bibitem{:2012sy} 
  D0 Collaboration, V.~M.~Abazov { et al.},
  \emph{Measurement of the ratio of three-jet to two-jet cross sections in $p\bar{p}$ collisions at $\sqrt{s}=1.96$ TeV,}
  [arXiv:1209.1140].

\bibitem{:2012xi} 
  D0 Collaboration, V.~M.~Abazov { et al.},
  \emph{Measurement of angular correlations of jets at sqrt(s)=1.96 TeV and determination of the strong coupling at high momentum transfers,}
  \emph{Phys. Lett. B} {\bf 718} (2012) 56
   [arXiv:1207.4957].

\bibitem{Abazov:2004hm} 
   D0 Collaboration, V.~M.~Abazov { et al.}, 
  \emph{Measurement of dijet azimuthal decorrelations at central rapidities in $p\bar{p}$ collisions at $\sqrt{s} = 1.96$ TeV,}
  \emph{Phys.\ Rev.\ Lett.}  {\bf 94}  (2005) 221801
  [hep-ex/0409040].

\bibitem{begel}
   M.~Begel, M.~Wobisch, and M.~Zielinski, 
  \emph{Dijet Azimuthal Decorrelations and Monte Carlo Tuning,}
  in \emph{Tevatron-for-LHC Report of the QCD Working Group,}
  FERMILAB-CONF-06-359,
  [hep-ph/0610012].

\bibitem{Khachatryan:2011zj} 
  CMS Collaboration, V.~Khachatryan { et al.},  
  \emph{Dijet Azimuthal Decorrelations in $pp$ Collisions at $\sqrt{s} = 7$~TeV,}
  \emph{Phys.\ Rev.\ Lett.}  {\bf 106}  (2011) 122003
  [arXiv:1101.5029].

\bibitem{daCosta:2011ni} 
  ATLAS Collaboration, G.~Aad { et al.},  
  \emph{Measurement of Dijet Azimuthal Decorrelations in pp Collisions 
   at $\sqrt{s}$ = 7 TeV,}
  \emph{Phys.\ Rev.\ Lett.}  {\bf 106} (2011) 172002
  [arXiv:1102.2696].

\bibitem{Dhullipudi}
   R.~Dhullipudi,
   \emph{The Study of $p_T$~Dependence of Dijet Azimuthal Decorrelations in 
   Proton-Proton Collisions at $\sqrt{s} = 7$~TeV,}
   Ph.D. dissertation, Louisiana Tech University (2012).

\bibitem{Chakravarthula} 
K.~Chakravarthula,
  \emph{Study of Jet Transverse Momentum and Jet Rapidity Dependence of Dijet Azimuthal 
  Decorrelations with the D0 Detector,}
  Ph.D. dissertation, Louisiana Tech University (2012).


\bibitem{Cacciari:2008gp} 
  M.~Cacciari, G.~P.~Salam, and G.~Soyez,
  \emph{The Anti-k(t) jet clustering algorithm,}
  \emph{JHEP} {\bf 0804}  (2008) 063
  [arXiv:0802.1189].


\bibitem{run2cone}  
  G.~C.~Blazey {et al.}, 
 \emph{Run II Jet Physics,}
 in:
  U.~Baur, R.~K.~Ellis, and D. Zeppenfeld (Eds.),
  {\sl Proceedings of the Workshop: QCD and Weak Boson
  Physics in Run II},    
  Fermilab-Pub-00/297 (2000).


\bibitem{Cacciari:2011ma} 
  M.~Cacciari, G.~P.~Salam, and G.~Soyez,
  \emph{FastJet User Manual,}
  \emph{Eur.\ Phys.\ J.\ C} {\bf 72} (2012) 1896
  [arXiv:1111.6097].


\bibitem{Cacciari:2005hq} 
  M.~Cacciari and G.~P.~Salam,
  \emph{Dispelling the $N^{3}$ myth for the $k_t$ jet-finder,}
  \emph{Phys.\ Lett.\ B} {\bf 641}  (2006) 57
  [hep-ph/0512210].


\bibitem{Nagy:2003tz} 
  Z.~Nagy,
  \emph{Next-to-leading order calculation of three jet observables 
   in hadron hadron collision,}
  \emph{Phys.\ Rev.\ D} {\bf 68}  (2003) 094002
  [hep-ph/0307268].


\bibitem{Nagy:2001fj} 
  Z.~Nagy,
  \emph{Three jet cross-sections in hadron hadron collisions at next-to-leading order,}
  \emph{Phys.\ Rev.\ Lett.}  {\bf 88}  (2002) 122003
  [hep-ph/0110315].


\bibitem{Kluge:2006xs} 
  T.~Kluge, K.~Rabbertz, and M.~Wobisch,
  \emph{FastNLO: Fast pQCD calculations for PDF fits,}
  in M. Kuze, K. Nagano, K. Tokushuku, K. Hackensack (Eds.),
  \emph{Proceedings of the XIV Workshop on Deep Inelastic Scattering,}
  [hep-ph/0609285].

\bibitem{Bardeen:1978yd}
  W.~A.~Bardeen {et al.},
  \emph{Deep Inelastic Scattering Beyond The Leading Order In Asymptotically Free
  Gauge Theories,}
  \emph{Phys.\ Rev.\  D} {\bf 18} (1978) 3998.


\bibitem{Martin:2009iq}
  A.~D.~Martin {et al.},
  \emph{Parton distributions for the LHC,}
  \emph{Eur.\ Phys.\ J.\  C} {\bf 63} (2009) 189
  [arXiv:0901.0002].

\bibitem{Lai:2010vv}
  H.~L.~Lai {et al.},
  \emph{New parton distributions for collider physics,}
  \emph{Phys.\ Rev.\  D} {\bf 82} (2010) 074024
  [arXiv:1007.2241].

\bibitem{Ball:2011mu}
  R.~D.~Ball {et al.},
  \emph{Impact of Heavy Quark Masses on Parton Distributions and LHC
  Phenomenology,}
  \emph{Nucl.\ Phys.\  B} {\bf 849} (2011) 296
   [arXiv:1101.1300].

\bibitem{Corcella:2000bw} 
  G.~Corcella {et al.},
  \emph{HERWIG 6: An Event generator for hadron emission reactions with interfering
 gluons (including supersymmetric processes),}
 \emph{JHEP} {\bf 0101} (2001) 010
  [hep-ph/0011363].

\bibitem{Corcella:2002jc} 
  G.~Corcella {et al.}, 
  \emph{HERWIG 6.5 release note,}
  [hep-ph/0210213].

\bibitem{pythia}
  T.~Sj\"ostrand {et al.},
  \emph{High-energy physics event generation with PYTHIA 6.1,}
   \emph{Comput.\ Phys.\ Commun.}  {\bf 135} (2001) 238
  [hep-ph/0010017].


\bibitem{Sjostrand:2006za} 
  T.~Sj\"ostrand, S.~Mrenna, and P.~Z.~Skands,
  \emph{PYTHIA 6.4 Physics and Manual,}
  \emph{JHEP} {\bf 0605}  (2006) 026
  [hep-ph/0603175].

\bibitem{Albrow:2006rt}
    R.~Field,
  \emph{Tevatron Run 2 Monte-Carlo Tunes}
  in \emph{Tevatron-for-LHC Report of the QCD Working Group,}
  FERMILAB-CONF-06-359,
  [hep-ph/0610012].

\bibitem{tuneA}
   R.~Field,
   \emph{Min-Bias and the Underlying Event at the Tevatron and the LHC,}
    talk presented at the Fermilab ME/MC Tuning Workshop, Fermilab, 
    October 4, 2002.

\bibitem{Diehl:2010zz}
  G. Brandt,
  \emph{Charged particle multiplicities in inelastic p p interactions with ATLAS,}
  in:
  M.~Diehl, J.~Haller, T.~Sch\"orner-Sadenius, and G.~Steinbr\"uck (Eds.),
  \emph{5th Conference: Physics at the LHC 2010},
  DESY-PROC-2010-01 (2010).


\bibitem{Schulz:2011qy} 
  H.~Schulz and P.~Z.~Skands,
  \emph{Energy Scaling of Minimum-Bias Tunes,}
  \emph{Eur.\ Phys.\ J.\ C} {\bf 71} (2011) 1644
  [arXiv:1103.3649].


\bibitem{Sjostrand:2004pf} 
  T.~Sj\"ostrand and P.~Z.~Skands,
  \emph{Multiple interactions and the structure of beam remnants,}
  \emph{JHEP} {\bf 0403}  (2004) 053
  [hep-ph/0402078].

\bibitem{Sjostrand:2004ef} 
  T.~Sj\"ostrand and P.~Z.~Skands,
  \emph{Transverse-momentum-ordered showers and interleaved multiple interactions,}
  \emph{Eur.\ Phys.\ J.\ C} {\bf 39} (2005) 129
  [hep-ph/0408302].


\bibitem{Abazov:2009nc} 
 D0 Collaboration,  V.~M.~Abazov {et al.},
  \emph{Determination of the strong coupling constant from the inclusive jet 
  cross section in $p\bar{p}$ collisions at sqrt(s)=1.96 TeV,}
  \emph{Phys.\ Rev.\ D} {\bf 80} (2009) 111107
  [arXiv:0911.2710].


\end{thebibliography}
\end{document}